\documentclass[preprint,showpacs,preprintnumbers,amsmath,amssymb]{rspublic}

\usepackage{graphicx}
\usepackage{dcolumn}
\usepackage{bm}

\newcommand{\nab}{{\bf{\nabla}}}

\newcommand{\lam}{\Lambda(t)}
\newcommand{\ps}{\Psi_\alpha}
\newcommand{\pa}{\partial}

\newcommand{\largex}{{\bf x}}
\newcommand{\largep}{{\bf p}}

\newcommand{\ep}{\varepsilon}

\begin{document}


\title[Fast-forward of adiabatic dynamics]{Fast-forward of adiabatic dynamics in
quantum mechanics}

\author[Shumpei Masuda \& Katsuhiro Nakamura]{Shumpei Masuda$^{1}$ and Katsuhiro
 Nakamura$^{2,3}$}

\affiliation{%
$^{1}$
Department of Physics,
Kwansei Gakuin University,
Gakuen, Sanda, Hyogo 669-1337, Japan\\
$^2$
Heat Physics Department,
Uzbek Academy of Sciences,
28 Katartal Str.,
100135 Tashkent,
Uzbekistan\\
$^3$
Department of Applied Physics,
Osaka City University,
Sumiyoshi-ku, Osaka 558-8585, Japan
}%

\date{\today}

\label{firstpage}
\maketitle
\begin{abstract}{atom manipulation, mechanical control of atoms, 
quantum transport}
We propose a way to accelerate adiabatic dynamics of wave functions
in quantum mechanics to
obtain a final adiabatic state 
except for the spatially uniform phase in any desired short time.
We develop the previous theory of fast-forward 
(Masuda \& Nakamura 2008)
so as to
derive a driving potential for the fast-forward of the adiabatic dynamics.
A typical example is
the fast-forward of adiabatic transport of 
a wave function which is the
ideal transport in the sense that a stationary wave function
is transported to an aimed position in any desired short time
without leaving any disturbance at the final time of the fast-forward.
As other 
important examples we show accelerated manipulations of wave functions 
such as their splitting and squeezing.
The theory is also applicable to macroscopic quantum mechanics described by the 
nonlinear Schr$\ddot{\mbox{o}}$dinger equation.
\end{abstract}


\section{Introduction}
\label{Introduction}
The adiabatic process occurs when the 
external parameter of Hamiltonian 
of the system is 
adiabatically changed.
Quantum adiabatic theorem (Born \& Fock 1928; Kato 1950; Messiah 1962), 
which states that, 
if the system is initially in an eigenstate of the instantaneous Hamiltonian, 
it remains so during the process,
has been studied in various contexts 
(Berry 1984; 
Aharonov \& Anandan 1987; 
Samuel \& Bhandari 1988; 
Shapere \& Wilczek 1989; 
Nakamura \& Rice 1994; 
Bouwmeester {\it et al.} 1996; 
Farhi {\it et al.} 2001; 
Roland \& Cerf 2002;
Sarandy \& Lidar 2005; 
Du {\it et al.} 2008).
The rate of a change in the parameter of Hamiltonian is infinitesimal,
so that it takes infinite time to achieve the final result in the process. 

In our previous paper (Masuda \& Nakamura 2008), we investigated 
a way to accelerate quantum dynamics with use of an
additional phase of a wave function (WF). 
We can accelerate a given quantum dynamics to obtain 
a target state in any desired short time.
This kind of acceleration is called the fast-forward of quantum dynamics.

The idea of the adiabatic process seems to be incompatible 
with that of fast-forward.
But here we combine these two ideas, that is,
we propose a theory to accelerate the adiabatic dynamics in quantum mechanics
and obtain, in any desired short time, the target state originally 
accessible after infinite time through the adiabatic dynamics.
By using this theory we can find a driving potential to generate
the target state exactly except for a spatially-uniform time-dependent phase 
like
dynamical and adiabatic phases (Berry 1984).
The fast-forward of the 
adiabatic dynamics makes an ideal transport of
quantum states possible:
a stationary wave packet is moved to an aimed position 
without leaving any disturbance
at the end of transport.
After the transport, the wave packet becomes stationary again, and is in
the same energy level as the initial one.

Before embarking upon the main part of the text,
we briefly summarize the previous theory of the fast-forward of quantum dynamics
(Masuda \& Nakamura 2008).
In Schr$\ddot{\mbox{o}}$dinger equation with a given potential 
$V_0=V_0(\largex,t)$ and nonlinearity constant $c_0$
(appearing in macroscopic quantum dynamics)
\begin{eqnarray}
i\hbar\frac{d \Psi_0}{dt}=-\frac{\hbar^2}{2m_0}\nab^2\Psi_0
+V_0(\largex,t)\Psi_0 - c_0|\Psi_0|^2\Psi_0,
\label{schr-1}
\end{eqnarray}
$\Psi_0(\largex,t)$ is supposed to be a known 
function of space ($\largex$) and time ($t$) and is called a  
standard state.
For any long time $T$ called a standard final time, we choose
$\Psi_0(t=T)$ as a target state that we are going to generate.
Let $\Psi_\alpha(\largex,t)$ be a virtually 
fast-forwarded state of $\Psi_0(\largex,t)$ defined by 
\begin{eqnarray}
|\Psi_\alpha(t)> = |\Psi_0(\alpha t)>,
\label{ps111}
\end{eqnarray}
where $\alpha$ is a time-independent magnification factor of the fast-forward.
The time-evolution of the WF is speeded up for $\alpha > 1$ and
slowed down for $0<\alpha<1$
like a slow-motion.
A rewind can occur for $\alpha<0$,
and the WF pauses when $\alpha=0$.

In general, the magnification factor can be time-dependent,
$\alpha = \alpha(t)$.
The time-evolution of a WF is accelerated and decelerated
when $\alpha(t)$ is increasing and decreasing, respectively.
In this case,
the virtually fast-forwarded state is defined as,
\begin{eqnarray}
  |\Psi_\alpha(t)> = |\Psi_0(\lam)>,
\label{exact_s}
\end{eqnarray}
where
\begin{eqnarray}
\lam = \int_0^{t}\alpha(t')dt'.
\label{lam}
\end{eqnarray}

Since the generation of $\ps$ requires an anomalous mass reduction,
we can not generate $\ps$ (Masuda \& Nakamura 2008). 
But we can obtain the target state by considering
a fast-forwarded state $\Psi_{FF} = \Psi_{FF}(\largex,t)$ which differs from
$\ps$ by an additional space-dependent phase,
$f=f(\largex, t)$, as
\begin{eqnarray}
\Psi_{FF}(t) = e^{if}\ps(t) = e^{if}\Psi_0(\lam).
\label{psiff0}
\end{eqnarray}
The Schr$\ddot{\mbox{o}}$dinger equation for $\Psi_{FF}$ is given by
\begin{eqnarray}
i\hbar\frac{d \Psi_{FF}}{dt}=-\frac{\hbar^2}{2m_0}\nab^2\Psi_{FF}
+V_{FF}(\largex,t)\Psi_{FF} - c_0|\Psi_{FF}|^2\Psi_{FF},
\label{schrff0}
\end{eqnarray}
where $V_{FF}(\largex,t)$ is called a driving potential.
If we appropriately 
tune the initial and final behaviours of the 
time dependence of $\alpha$ (the detail is shown later), 
the additional phase can vanish
at the final time of the fast-forward $T_{F}$, and we can obtain the 
exact target
state, that is,   
\begin{eqnarray}
\Psi_{FF}(T_F) = \Psi_0(T),
\end{eqnarray}
where $T_F$ is the final time of the fast-forward defined by
\begin{eqnarray}
T=\int_0^{T_F}\alpha(t)dt.\label{tf1}
\end{eqnarray}
(In the case of a constant $\alpha$, $T_{F}=T/\alpha$.)

From equations (\ref{schr-1}), (\ref{lam}), (\ref{psiff0})
and (\ref{schrff0}) we obtain
 an equation for the additional phase $f$ (Masuda \& Nakamura 2008)
\begin{eqnarray}
&&|\Psi_\alpha|^2{\bf\nabla}^2f+
2\mbox{Re}[\Psi_\alpha\nabla\Psi_\alpha^\ast]\cdot {\bf\nabla}f\nonumber\\
&&\hspace{2cm} +(\alpha-1) 
\mbox{Im} [\Psi_\alpha\nabla^2\Psi_\alpha^\ast] = 0,
\label{feq44_2}
\end{eqnarray}
and the driving potential of the fast-forward $V_{FF}$
\begin{eqnarray}
V_{FF} &=& \alpha V_0  - \hbar\frac{d f}{d t} 
- \frac{\hbar^2}{2m_0}({\bf\nabla}f)^2\nonumber\\ 
&&+ \mbox{Re}[ -(\alpha-1)\frac{\hbar^2}{2m_0} 
{\bf\nabla}^2 \Psi_\alpha/\Psi_\alpha
+ i\frac{\hbar^2}{m_0}{\bf\nabla}f\cdot {{\bf\nabla}\Psi_\alpha}/{\ps}]
\nonumber\\ 
&&-(\alpha-1)c_0|\Psi_\alpha|^2.
\label{veq22}
\end{eqnarray}
With use of the phase $\eta=\eta(\largex,t)$ of the standard state $\Psi_0$,
$f$ is given by (Masuda \& Nakamura 2008)
\begin{eqnarray}
f(\largex, t) = (\alpha(t)-1) \eta(\largex,\lam),
\label{f0}
\end{eqnarray}
which satisfies equation (\ref{feq44_2}) and determines $V_{FF}$ in equation 
(\ref{veq22}).
We impose the initial and final conditions for $\alpha(t)$:
$\alpha$ must start from $1$, increase for a while
and decrease back to $1$ with $d\alpha/dt = 0$ at the final time of the 
fast-forward. Then the additional phase $f$ vanishes in the initial and final
time of the fast-forward.
Once we have a standard state,
we can obtain a target state in any desired short time by applying the driving
potential with suitably tuned $\alpha(t)$.

However, in the fast-forward of the adiabatic dynamics
we shall use infinitely-large $\alpha$.
Then the expression of $V_{FF}$ in equation (\ref{veq22}) and $f$ in 
equation (\ref{f0}) 
should diverge.
This difficulty will be overcome by regularization of the standard potential 
and states which will be described in Section
\ref{Wave function under adiabatic dynamics}.
In Section \ref{examples} we apply the theory
to the ideal transport
by accelerating the adiabatic transport of a wave function in
a moving confining potential.
Ideal manipulations 
such as wave packet 
splitting and squeezing are shown as other examples
of the
fast-forward of adiabatic dynamics.
Section \ref{Conclusion} is devoted to summary and discussions.

\section{Regularization of standard state and driving potential for fast-forward}
\label{Wave function under adiabatic dynamics}
\subsection{Difficulty involved in standard adiabatic dynamics}
\label{Some problems involved in standard dynamics}
We consider a dynamics of a WF, $\Psi_0$, under the potential:
$V_0 = V_0(\largex,R(t))$ which is adiabatically varies,
where $R=R(t)$ is a parameter in the potential which is adiabatically changed 
from a constant $R_0$ as
\begin{eqnarray}
R(t) = R_0 + \ep t.
\label{R}
\end{eqnarray}
The constant value $\ep$ is the rate of adiabatic change of $R(t)$ 
with respect to time
and is infinitesimal, that is,
\begin{eqnarray}
\frac{d R(t)}{d t} &=& \varepsilon,\label{dRdt}\\
\ep &\ll& 1.
\end{eqnarray}
The Hamiltonian is represented as
\begin{eqnarray}
H_0 = \frac{\largep^2}{2m_0} + V_0(\largex,R(t)),
\label{H}
\end{eqnarray}
and Schr$\ddot{\mbox{o}}$dinger equation for $\Psi_0$ is given as
\begin{eqnarray}
i\hbar\frac{d \Psi_0}{dt}=-\frac{\hbar^2}{2m_0}\nab^2\Psi_0
+V_0(\largex,R(t))\Psi_0 - c_0|\Psi_0|^2\Psi_0,
\label{schr1}
\end{eqnarray}
where $c_0$ is a nonlinearity constant.
If a system is in the $n$-th energy eigenstate at the 
initial time, the adiabatic 
theorem guarantees that, in the limit $\ep\rightarrow 0$, 
$\Psi_0$ remains in the $n$-th energy eigenstate of 
the instantaneous Hamiltonian.
Then $\Psi_0$ is represented as
\begin{eqnarray}
\Psi_0(\largex,t,R(t)) = \phi_n(\largex,R(t)) 
e^{-\frac{i}{\hbar}\int_0^t E_n(R(t'))dt'}
e^{i\Gamma(t)},
\label{adia1}
\end{eqnarray}
where $E_n=E_n(R)$ and 
$\phi_n=\phi_n(\largex,R)$ are the $n$-th energy eigenvalue and
eigenstate corresponding to the parameter
$R$, respectively,
and $\Gamma = \Gamma(t)$ is the adiabatic phase defined by
\begin{eqnarray}
\Gamma(t) = i\int_{0}^t\int_{-\infty}^\infty d\largex dt
\phi_n^\ast\frac{d}{dt}\phi_n,
\label{gamma}
\end{eqnarray}
which is independent of space coordinates.
$\phi_n$ fulfills
\begin{eqnarray}
\frac{\pa \phi_n}{\pa t} &=& 0,\label{dphidt}\\
-\frac{\hbar^2}{2m_0}\nab^2\phi_n + V_0(\largex,R)\phi_n &-& c_0|\phi_n|^2\phi_n 
= E_n(R)\phi_n\label{tis}.
\end{eqnarray}
The second factor in the right hand side of equation (\ref{adia1})
is called a dynamical phase factor,
which is also space-independent.
Such ideal adiabatic dynamics with $\ep\rightarrow 0$ takes infinite time
until we have an aimed adiabatic state (target state).

Our aim is to accelerate an adiabatic dynamics 
$\phi_n(R(0))\rightarrow \phi_n(R(T))$ 
aside from the dynamical and adiabatic phases
by applying the theory of 
fast-forward with infinitely-large magnification factor $\alpha$, and obtain
a target state $\phi_n(R(T))$ in any desired short time, 
where $T$ is a standard final time which is taken
to be $O(1/\ep)$.
During the fast-forward, we will take the limit $\ep\rightarrow 0$, 
$\alpha\rightarrow\infty$, $T\rightarrow\infty$ and $\alpha\ep \sim 1$.
In this acceleration, we do not care about spatially uniform phase.
(The spatially uniform phase can be controlled by spatially uniform
potential, if it is necessary.)
In applying the fast-forward to the adiabatic dynamics, 
we have to choose a standard state and Hamiltonian.
But there is some ambiguity in the choice of the standard state and Hamiltonian,
because the state which we want to accelerate is represented in the limit
$\ep\rightarrow 0$ and we will fast-forward it with infinitely-large 
magnification factor. 
One might think that we can take $\Psi_0$ and $H_0$ 
in equations (\ref{adia1}) and (\ref{H}) as a standard state and Hamiltonian.
However, such idea is not adequate
because the state in equation (\ref{adia1}) is in an expression of WF 
in the limit $\ep\rightarrow 0$ and does not 
satisfy Schr$\ddot{\mbox{o}}$dinger equation
up to $O(\ep)$ with small but finite $\ep$ (Kato 1950; Wu \& Yang 2005). 
In other words quantum dynamics in equation (\ref{schr-1}) with finite $\ep$
inevitably induces nonadiabatic transition, but $\Psi_0$ in equation
(\ref{adia1})
ignores such transitions.
Then the original theory of fast-forward 
(Masuda \& Nakamura 2008) is not applicable as it stands.
To overcome this difficulty, we shall regularize the
standard state and Hamiltonian corresponding to
the adiabatic dynamics.

In the fast-forward of the adiabatic dynamics
in the limit $\ep\rightarrow 0$, 
$\alpha\rightarrow\infty$ and $\alpha\ep \sim 1$,
the standard final time $T$ is chosen as $T = O(\frac{1}{\ep})$.
In this case, a regularized standard state and Hamiltonian
should fulfill the following conditions:\\
$1$. A regularized 
standard Hamiltonian and state of the fast-forward should agree
with  $H_0$ and $\Psi_0$ 
except for 
space-independent phase, respectively, in the limit $\ep\rightarrow 0$;\\
$2$. The regularized standard state should satisfy 
the Schr$\ddot{\mbox{o}}$dinger equation corresponding 
to the regularized standard Hamiltonian up to $O(\ep)$ with finite $\ep$.

Hereafter $\Psi_0^{(reg)}$ and $H_0^{(reg)}$ are prescribed to
a regularized standard state and Hamiltonian, respectively, which
fulfill the conditions $1$ and $2$.

\subsection{Regularized standard state}
Here we derive expressions for $\Psi_0^{(reg)}$ and $H_0^{(reg)}$.
Let us consider a regularized Hamiltonian $H_0^{(reg)}$: 
\begin{eqnarray}
H_0^{(reg)} = \frac{\largep^2}{2m_0} + V_0^{(reg)},
\label{H0}
\end{eqnarray}
where the potential $V_0^{(reg)}$ in the regularized
Hamiltonian,
is given as
\begin{eqnarray}
V_0^{(reg)}(\largex,t) = V_0(\largex,R(t)) + \varepsilon \tilde{V}(\largex,t).
\label{sp1}
\end{eqnarray}
$\tilde{V}$ is a real function of $\largex$ and $t$ to be determined 
{\it a posteriori}, which is introduced to incorporate the effect of 
non-adiabatic transitions.
It is obvious that 
$H_0^{(reg)}$ agrees with $H_0$ in the limit $\ep \rightarrow 0$, that is,
\begin{eqnarray}
\lim_{\ep\rightarrow 0}H_0^{(reg)}(\largex,t) = H_0(\largex,R(t)).
\end{eqnarray}
Suppose the regularized standard state to be represented as
\begin{eqnarray}
\Psi_0^{(reg)}(\largex,t,R(t)) = \phi_n(\largex,R(t)) 
e^{-\frac{i}{\hbar}\int_0^t E_n(R(t'))dt'}
e^{i\varepsilon\theta(\largex,t)},
\label{standard1}
\end{eqnarray}
where $\theta$ is a real function of $\largex$ and $t$. 
$\phi_n$ and $E_n$ are the same as used in equation (\ref{adia1}).
$\Psi_0^{(reg)}$ coincides with
$\Psi_0$ except for the spatially uniform phase 
in the limit $\ep \rightarrow 0$.

Thus $\Psi_0^{(reg)}$ in equation (\ref{standard1}) and $H_0^{(reg)}$ in
equation (\ref{H0}) fulfill
the condition $1$ given in the previous subsection. 
$\theta$ and $\tilde{V}$ are chosen so that 
$\Psi_0^{(reg)}$ satisfies Schr$\ddot{\mbox{o}}$dinger equation 
corresponding to $H_0^{(reg)}$ up to $O(\ep)$ to fulfill the condition $2$.

Here we shall obtain $\theta$ and $\tilde{V}$.
Schr$\ddot{\mbox{o}}$dinger equation for $\Psi_0^{(reg)}$ is represented as
\begin{eqnarray}
i\hbar\frac{d \Psi_0^{(reg)}}{dt}=-\frac{\hbar^2}{2m_0}\nab^2\Psi_0^{(reg)}
+V_0^{(reg)}(\largex,R(t))\Psi_0^{(reg)} - c_0|\Psi_0^{(reg)}|^2\Psi_0^{(reg)}.
\label{schr1_1}
\end{eqnarray}
Substituting equations (\ref{sp1}) and (\ref{standard1}) into 
equation (\ref{schr1_1}),
we find
\begin{eqnarray}
i\hbar\frac{\pa\phi_n}{\pa R}\varepsilon + E_n\phi_n
- \hbar\frac{d\theta}{d t}\varepsilon\phi_n
&=& -\frac{\hbar^2}{2m_0}[\nab^2\phi_n + 2i\varepsilon\nab\theta\cdot\nab\phi_n
-\varepsilon^2(\nab\theta)^2\phi_n \nonumber\\ 
&&+ i\varepsilon(\nab^2\theta)\phi_n]
+ V\phi_n + \varepsilon \tilde{V}\phi_n \nonumber\\  
&&- c_0|\phi_n|^2\phi_n.
\label{schr2}
\end{eqnarray}
With use of equation (\ref{tis}) in equation (\ref{schr2}), we have
\begin{eqnarray}
i\hbar\frac{\pa\phi_n}{\pa R}
- \hbar\frac{d\theta}{d t}\phi_n
=-\frac{\hbar^2}{2m_0}[2i\nab\theta\cdot\nab\phi_n
+i(\nab^2\theta)\phi_n] + \tilde{V}\phi_n,\label{schr3}
\end{eqnarray}
where we eliminated the term of the second order in $\ep$.
Multiplying both sides of equation 
(\ref{schr3}) by $\frac{i}{\hbar}\phi_n^\ast$, 
we have
\begin{eqnarray}
-\phi_n^\ast\frac{\pa\phi_n}{\pa R}
-i\frac{d\theta}{dt}|\phi_n|^2
=\frac{\hbar}{2m_0}[2\phi_n^\ast\nab\phi_n\cdot\nab\theta +
|\phi_n|^2\nab^2\theta] + i\frac{\tilde{V}}{\hbar}|\phi_n|^2.\label{schr4}
\end{eqnarray}
Equation 
(\ref{schr4}) is equivalently represented by decomposing it into real and 
imaginary parts as 
\begin{eqnarray}
|\phi_n|^2\nab^2\theta + 2\mbox{Re}[\phi_n\nab\phi_n^\ast]\cdot\nab\theta
+\frac{2m_0}{\hbar}\mbox{Re}[\phi_n\frac{\pa\phi_n^\ast}{\pa R}] = 0,
\label{etheta2}\\
\frac{\hbar}{m_0}\mbox{Im}[\phi_n^\ast\nab\phi_n]\cdot\nab\theta + 
\frac{\tilde{V}}{\hbar}|\phi_n|^2+\mbox{Im}[\phi_n^\ast\frac{\pa}{\pa R}\phi_n]
+ \frac{d\theta}{dt}|\phi_n^2|= 0.\label{imag1}
\end{eqnarray}
$\theta(\largex,t)$ should satisfy equation (\ref{etheta2}).
Equation (\ref{etheta2}) can be derived also from the continuity equation for 
$\Psi_0^{(reg)}$
(See \ref{Approach from the continuity equation}).
$\tilde{V}$ is then given in terms of $\theta$ and $\phi_n$, that is,
\begin{eqnarray}
\tilde{V} = -\hbar\frac{d\theta}{d t} - \hbar\mbox{Im}
[\frac{\pa \phi_n}{\pa R}/\phi_n]  - 
\frac{\hbar^2}{m_0}\mbox{Im}[\frac{\nab\phi_n}{\phi_n}]\cdot\nab\theta.
\label{V'0}
\end{eqnarray}
Equation (\ref{etheta2}) indicates that $\theta$ is a function which is
not dependent on $t$ explicitly, leading to
\begin{eqnarray}
\frac{d\theta}{d t} = \frac{\pa R}{\pa t}\frac{\pa\theta}{\pa R}
=\ep \frac{\pa\theta}{\pa R}.
\end{eqnarray}
Therefore, in our approximation to suppress the terms of $O(\ep^2)$, 
equation (\ref{V'0}) is reduced to
\begin{eqnarray}
\tilde{V} = - \hbar\mbox{Im}
[\frac{\pa \phi_n}{\pa R}/\phi_n] - 
\frac{\hbar^2}{m_0}\mbox{Im}[\frac{\nab\phi_n}{\phi_n}]\cdot\nab\theta.
\label{V'3}
\end{eqnarray}
Thus we have obtained equations which $\theta$ and $\tilde{V}$ should satisfy
so that the conditions of regularized standard state and Hamiltonian 
are fulfilled.


\subsection{Driving potential for fast-forward}
\label{Driving potential for Fast-forward of adiabatic dynamics}
Here we obtain the driving potential $V_{FF}$ to fast-forward 
the regularized standard state $\Psi_0^{(reg)}$. 
Such limit are taken as $\ep \rightarrow 0$, 
$\alpha \rightarrow \infty$ and $\alpha\ep \sim 1$ in the fast-forward.
The final time of the standard dynamics $T$ is taken to be $O(1/\ep)$.  
The driving potential $V_{FF}=V_{FF}(\largex,t)$ in equation (\ref{veq22})
is explicitly represented as
\begin{eqnarray}
V_{FF}(\largex,t) &=& \alpha V_0^{(reg)}(\largex,R(\lam)) - \hbar\frac{d f}{d t}
(\largex,t)
-\frac{\hbar^2}{2m_0}(\nab f)^2 \nonumber\\ 
&& + \mbox{Re}[-(\alpha-1)\frac{\hbar^2}{2m_0}\nab^2\Psi_0^{(reg)}(\largex,\lam)/\Psi_0^{(reg)}  \nonumber\\ 
&& + i\frac{\hbar^2}{m_0} \nab f\cdot\nab\Psi_0^{(reg)}/\Psi_0^{(reg)}]
 - (\alpha-1)c_0|\Psi_0^{(reg)}|^2,
\label{vff0_2}
\end{eqnarray}
where $f=f(\largex,t)$ is the additional phase of the fast-forwarded state 
$\Psi_{FF}$:
\begin{eqnarray}
\Psi_{FF}(\largex,t) = 
e^{if(\largex,t)}\Psi_0^{(reg)}(\largex,\Lambda(t))
\end{eqnarray}
with $\lam$ defined by equation (\ref{lam}).

As described
in Section \ref{Introduction}, 
the additional phase $f$ given by equation (\ref{f0}) 
is not convenient
for the fast-forward of adiabatic dynamics, 
because $V_{FF}$ in equation (\ref{vff0_2}) 
would diverge due to the infinitely-large $\alpha$.
Instead we can take $\nab f$ as 
\begin{eqnarray}
\nab f=(\alpha-1)\varepsilon\nab\theta,\label{nabf}
\end{eqnarray}
which also satisfies equation (\ref{feq44_2}) 
(see \ref{Additional phase for fast-forward state}).
So we can determine $f$ such that
\begin{eqnarray}
f(\largex,t)=(\alpha(t)-1)\varepsilon\theta(\largex,\lam).\label{f1}
\end{eqnarray}
With use of $f$ in equation (\ref{f1}), we can avoid the divergence of the
driving potential.
By using equations (\ref{tis}), (\ref{sp1}), (\ref{standard1}) and (\ref{f1})
in equation (\ref{vff0_2}), 
the driving potential $V_{FF}$ is written as
\begin{eqnarray}
V_{FF} &=&  \alpha\ep \tilde{V} 
+ V_0 -(\alpha-1)\frac{\hbar^2}{2m_0}\varepsilon^2(\nab\theta)^2\nonumber\\
&& -\hbar\frac{d \alpha}{d t}\varepsilon\theta 
-\hbar(\alpha-1)\varepsilon\frac{d \theta}{d t}\nonumber\\
&&- \frac{\hbar^2}{2m_0}(\alpha-1)^2\varepsilon^2(\nab \theta)^2+ 
(\alpha-1)E_n(R(\Lambda)).\label{vff2}
\end{eqnarray}
Noting $\frac{d \theta(R(\lam))}{d t}= 
\alpha(t)\varepsilon\frac{\pa\theta}{\pa R}$,
in the limit $\varepsilon\rightarrow 0$,
$\alpha\rightarrow \infty$, 
and $\varepsilon\alpha\sim 1$
we can suppress the terms of $O(\alpha^p\ep^q)$ with $q>p\ge 0$ 
in equation (\ref{vff2}).
While the last term in equation (\ref{vff2}) would diverge, we omit it because
it concerns only with space-independent phase of $\Psi_{FF}$
and has no effect on the dynamics governed by $V_{FF}$.
Then 
the driving potential reduces to
\begin{eqnarray}
V_{FF} = \alpha\ep \tilde{V} + V_0 - \hbar\frac{d\alpha}{d t}\varepsilon\theta
-\hbar\alpha^2\varepsilon^2\frac{\pa \theta}{\pa R} 
- \frac{\hbar^2}{2m_0}\alpha^2
\varepsilon^2(\nab\theta)^2.
\label{Vff0}
\end{eqnarray}

From equation (\ref{Vff0}) the driving potential coincides with
the standard potential $V_0$ when $\alpha\ep=0$ and $d\alpha/dt = 0$.
Thus, 
$\alpha\ep$ should grow 
from $0$ and come back to $0$ at $t=T_F$,
where $T_F$ is the final time of the fast-forward defined by equation 
(\ref{tf1}).
On the other hand, $\alpha\ep = O(1)$ during the fast-forward. 
The change in the parameter $R$ in equation (\ref{R}), 
during $t=0$ and $T_{F}$, is given by
\begin{eqnarray}
\Delta R \equiv R(\Lambda(T_F))-R(0) = \ep\int_0^{T_F}\alpha(t)dt.
\end{eqnarray}
What we have to do for the fast-forward of an adiabatic dynamics is to obtain 
$\theta$ from a given adiabatic dynamics $\phi_n(R(t))$
and then apply the driving potential $V_{FF}$ given in terms of 
$\theta$ and $\phi_n$ in equation (\ref{Vff0}).

\section{Examples}
\label{examples}
We show some examples of the fast-forward of adiabatic processes
in the linear regime ($c_0 = 0$),
by numerically iterating equation (\ref{schrff0}) 
with $V_{FF}(\largex,t)$ in equation (\ref{Vff0}).
(Examples in the nonlinear regime ($c_0\ne 0$) will be reported elsewhere.)
We can obtain the target state in the 
adiabatic process in any 
desired short time by applying the theory in Section 
\ref{Wave function under adiabatic dynamics}.
In the following examples,
the magnification factor is commonly chosen for $0\le t \le T_F$ as  
\begin{eqnarray}
\alpha(t)\ep = \bar{v}\cos(\frac{2\pi}{T_F}*t + \pi) + \bar{v},
\label{valpha1}
\end{eqnarray}
where $\bar{v}$ is time average of $\alpha(t)\ep$ during the fast-forwarding, 
and the final time of the fast-forward
$T_F$ is related to the standard final time $T$ as $T_F=\ep T/\bar{v}$ 
(see equation (\ref{tf1})).
\begin{figure}[h]
\begin{center}
\includegraphics[width=7cm]{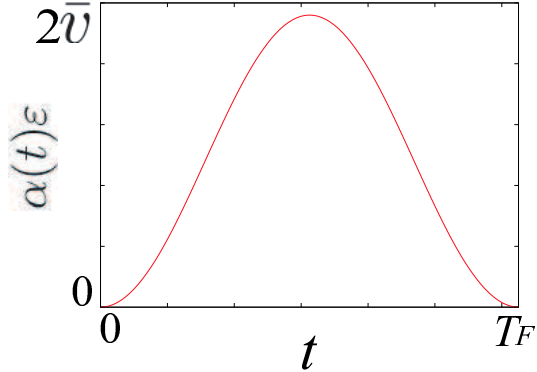}
\end{center}
\caption{\label{fig:epsart}
Time dependence of $\alpha(t)\ep$ for $0\le t \le T_F$. 
}
\label{valpha}
\end{figure}
We take $\ep T$ and $T_F$ as finite, where $\ep$ is infinitesimal and $T$ is
infinitely-large. Namely we aim to obtain the target state in finite time,
while the state is supposed to be obtained after infinitely long time $T$
in the original adiabatic dynamics. 
The time dependence of $\alpha(t)\ep$ is shown in figure \ref{valpha}.

\subsection{Fast-forward of adiabatic transport}
\label{Fast-forward of adiabatic transport}
Suppose we have a stationary wave packet (WP) in a confining potential 
in 2 dimensions and 
we want to transport the WP without leaving any disturbance on the wave function
after the transport.
If we move the confining potential rapidly to transport the WP fast,
the WF 
is radically affected, 
and it will oscillate in the confining 
potential after the transport.
However we can realize an ideal transport without leaving 
any disturbance at the end of
the transport by applying the present theory of
fast-forward of the adiabatic transport.
Suppose there is a stationary WF, $\psi(\largex)e^{-\frac{i}{\hbar}E_nt}$,
in the confining potential $U=U(x,y)$.
Then let the potential adiabatically slide with infinitesimal velocity $\ep$
in $x$-direction.
The sliding potential $V_0=V_0(x,y,t)$ is given by
\begin{eqnarray}
V_0 = U(x -\ep t,y).
\end{eqnarray}
The regularized WF in the potential is supposed to be given as
\begin{eqnarray}
\Psi_0^{(reg)} = \phi_ne^{-\frac{i}{\hbar}E_nt}e^{i\ep\theta},
\label{Psireg}
\end{eqnarray}
where
\begin{eqnarray}
\phi_n  &=& \psi(x -\ep t,y) = \psi(x-R(t),y),\label{phin2}\\
R(t) &=& \ep t.
\end{eqnarray}
From equation (\ref{phin2}) we have
\begin{eqnarray}
\frac{\pa \phi_n}{\pa R} = -\frac{\pa \phi_n}{\pa x}.
\label{phin3}
\end{eqnarray}
Substituting equation (\ref{phin3}) into equation (\ref{etheta2}) or 
equation (\ref{etheta0}) 
we have
\begin{eqnarray}
 \frac{\hbar}{m_0}\nab\cdot[|\phi_n|^2\nab\theta]
=2\mbox{Re}[\frac{\pa\phi_n^\ast}{\pa x}\phi_n].
\label{etheta3}
\end{eqnarray}
It can be easily confirmed that  
\begin{eqnarray}
\theta &=& \frac{m_0}{\hbar}x\label{theta2}
\end{eqnarray}
fulfill equation (\ref{etheta3}) for any function $\phi_n$.
By substituting equations (\ref{phin3}) and (\ref{theta2}) 
into equation (\ref{V'3})
we find
\begin{eqnarray}
 \tilde{V} = 0,
\label{vtilde1}
\end{eqnarray}
and the potential $V_0^{(reg)}(x,y,t)$ in equation (\ref{sp1}) 
in the regularized Hamiltonian
$H^{(reg)}_0$ is given as
\begin{eqnarray}
V_0^{(reg)} = V_0 = U(x -\ep t,y).
\label{V'2}
\end{eqnarray}
It should be noted that we can also derive equations 
(\ref{theta2}) and (\ref{vtilde1}) 
from the Galilean transformation of the coordinates
and omitting phase of WF in $O(\ep^2)$ 
which does not affect the fast-forward.

With use of equations (\ref{theta2}) and (\ref{V'2}) in equation (\ref{Vff0}),
we have
\begin{eqnarray}
V_{FF}(x,y,t) = U(x -\ep \Lambda(t),y) 
- \frac{d\alpha}{dt}m_0\ep x -\frac{m_0}{2}\alpha^2\ep^2.
\label{vff3}
\end{eqnarray}
Since we have derived $\theta$ without giving any specific form of $\phi_n$, 
the formula of the driving potential in equation (\ref{vff3}) 
is independent of the profile of 
the WF that we are going to transport.

As an example of the fast-forward of the adiabatic transport, we 
choose a ground state in a harmonic potential: 
$V_0^{(reg)}=\frac{m_0\omega^2}{2}\{(x-x_0(t))^2+(y-y_0)^2\}$. 
The centre of potential is adiabatically moved in $x$-direction as
\begin{eqnarray}
x_0(t) = x_0(0) + \epsilon t.
\end{eqnarray} 
$x_0$ corresponds to $R$ in Section
\ref{Wave function under adiabatic dynamics}.
The regularized standard state $\Psi_0^{(reg)}$ is
\begin{eqnarray}
\Psi_0^{(reg)}(\largex,t) = \phi_ne^{i\ep\theta} &\equiv& 
(\frac{m_0\omega}{\pi\hbar})^{\frac{1}{2}}
\exp[-\frac{m_0\omega}{2\hbar}\{(x-x_0(t))^2+(y-y_0)^2\}\nonumber \\
&&-i\omega t]e^{i\ep\theta},
\end{eqnarray}
with $\theta$ given by equation (\ref{theta2}).
We put $\ep T = 20$ and $T_F = 6.25$, where $\ep$ is infinitesimal and $T$ is
infinitely-large. 
The centre of mass moves from $(x,y)=(x_0(0),y_0)$
to $(x_0(0)+\ep T,y_0)$ during the fast-forward, where $y_0$ is a constant.
\begin{figure}[h]
\begin{center}
\includegraphics[width=7cm]{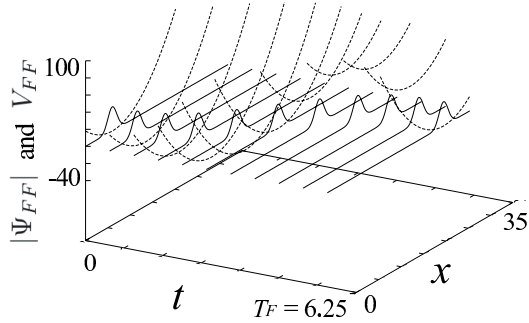}
\end{center}
\caption{\label{fig:epsart}
Spatio-temporal dependence of the driving potential (broken line) and
the amplitude of the fast-forwarded state $|\Psi_{FF}|^2$ (solid line) at 
$y=y_0$.
The parameters are taken as $\ep T = 20$, $T_F = 6.25$, $\frac{\hbar}{m_0}=1$,
$\omega= 1$, $x_0(0)=6$ and $y_0=16$.
}
\label{slid_p_v}
\end{figure}
Figure \ref{slid_p_v} is the 
spatio-temporal dependence of the driving potential (broken line) and
the amplitude of the fast-forwarded state (solid line) at $y=y_0$.
\begin{figure}[h]
\begin{center}
\includegraphics[width=7cm]{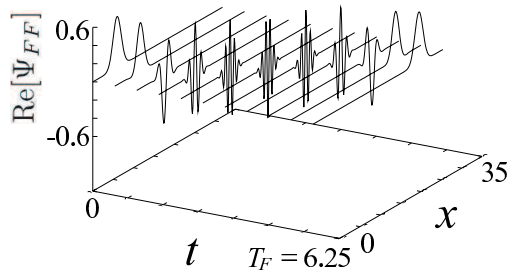}
\end{center}
\caption{\label{fig:epsart}
Spatio-temporal dependence of the real part of wave function of 
the fast-forwarded state at $y=y_0$. 
The parameters are the same as in figure \ref{slid_p_v}.
}
\label{slid_real}
\end{figure}
Figure \ref{slid_real} is spatio-temporal dependence 
of the real part of WF of 
the fast-forwarded state at $y=y_0$. 
Due to the additional phase $f$, the real part of WF shows spatial oscillations
during the fast-forward, but at the final time of the fast-forward such an 
oscillation disappears and $\Psi_{FF}$ agrees with the target state except for
the spatially uniform phase.
We can also fast-forward the adiabatic transport of WP 
in any excited state,
and obtain the target state in finite time.
This is the ideal transport of WPs because we can obtain the target 
state without any disturbances at the final time of the fast-forward.
And WP becomes stationary again after the transport when $\alpha\ep=0$ and 
$\frac{d\alpha}{dt}=0$.

\begin{figure}[h]
\begin{center}
\includegraphics[width=7cm]{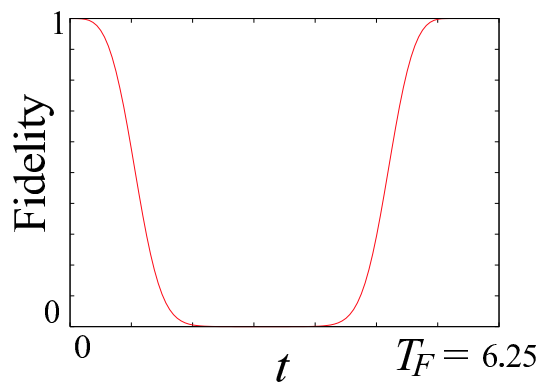}
\end{center}
\caption{\label{fig:epsart}
The time dependence of the fidelity.
}
\label{fidelity_transport}
\end{figure}

To numerically check the validity of the fast-forward, we calculate the fidelity
defined as
\begin{eqnarray}
F = |<\Psi_{FF}(t)|\Psi_0^{(reg)}(\lam)>|,
\label{fidelity}
\end{eqnarray}
i.e. the overlap between the fast-forwarded state $\Psi_{FF}(t)$
and the corresponding standard one $\Psi_0^{(reg)}(\lam)$,
where $\Psi_0^{(reg)}(\Lambda(T_F))\equiv \Psi_0^{(reg)}(T)$.
The time dependence of the fidelity defined by equation 
(\ref{fidelity}) is shown in
figure \ref{fidelity_transport}.
We find that the fidelity 
first decreases from unity due to the additional phase $f$ 
of the fast-forwarded state,
but at the final time it becomes unity with high precision ($|1-F|\le 10^{-5}$) 
again.
We have thus 
obtained the adiabatically-accessible 
target state in a good accuracy in finite time $T_F=6.25$.

\subsection{Fast-forward of wave packet splitting}
\begin{figure}[h]
\begin{center}
\includegraphics[width=7cm]{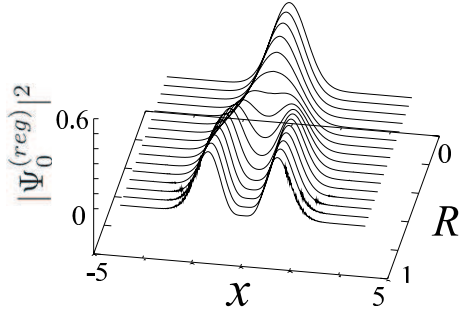}
\end{center}
\caption{\label{fig:epsart}
$x$-dependence of $|\Psi_0^{(reg)}|^2$ for various $R$.
}
\label{WF_split}
\end{figure}

\begin{figure}[h]
\begin{center}
\includegraphics[width=7cm]{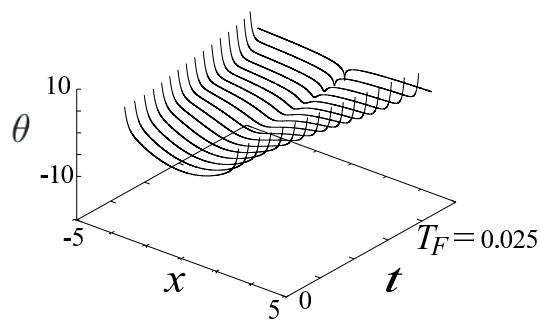}
\end{center}
\caption{\label{fig:epsart}
Spatio-temporal dependence of $\theta$.  
The parameters are chosen as $\ep T=1$, $T_F=0.025$, $\frac{\hbar}{m_0}=1.0$ and
$a = 1$.
}
\label{theta4}
\end{figure}

We then show the fast-forward of WP splitting in 1-dimension.
Here we calculate $\theta$ by a numerical integration of equation 
(\ref{etheta2})
or equation (\ref{etheta0}).
Let us consider the dynamics of a Gaussian wave packet (WP)
to be split into a pair of separated WPs.
The regularized standard WF is chosen as
\begin{eqnarray}
\Psi_0^{(reg)} = \phi_ne^{i\ep\theta} \equiv
h(R)\{(1-R)\exp[-\frac{a}{2}x^2] + Rx^2\exp[-\frac{a}{2}x^2]\}
e^{i\ep\theta},
\label{psi0_split}
\end{eqnarray}
where $R$ is defined by equation (\ref{R}) with $R_0=0$,
and $h(R)$ is a normalization constant represented as
\begin{eqnarray}
h(R) = [(1-R)^2\sqrt{\frac{\pi}{a}} + (1-R)R\frac{\sqrt{\pi}}{a^{3/2}}
+\frac{3}{4}R^2\frac{\sqrt{\pi}}{a^{5/2}}]^{-1/2}.
\end{eqnarray}
The potential in the regularized standard Hamiltonian is represented as
\begin{eqnarray}
V_0^{(reg)} = \frac{\hbar^2}{2m_0}\frac{Ra^2x^4 + \{(1-R)a^2-5Ra\}x^2+
2R-(1-R)a}{Rx^2+(1-R)}.
\label{V0_split}
\end{eqnarray}
Note that $\tilde{V}$ in equation (\ref{V'3}) 
vanishes due to the absence of space-dependent phase of 
$O(1)$ 
($\mbox{Im}(\frac{\pa\phi_n}{\pa R}/\phi_n)
=\mbox{Im}(\frac{\pa\phi_n}{\pa x}/\phi_n)=0$)
in the regularized standard state in equation (\ref{psi0_split}).
It is easily confirmed that for constant $R$, $\Psi_0^{(reg)}$ in 
equation (\ref{psi0_split}) with
$\theta=0$ is a zero-energy eigenstate under the potential $V_0^{(reg)}$ 
in equation (\ref{V0_split}).
While for $R=0$ the WF is a simple Gaussian, 
for $R=1$ it becomes spatially-separated double Gaussians.
The spatial distribution of $|\Psi_0^{(reg)}|^2$ for various $R$ from $0$ to $1$ 
is shown in figure \ref{WF_split}.
The standard final time $T$ is taken as $T = 1/\ep$ and $R(0)=0$.
As $R$ is gradually changed from $0$ to $1$,
$\Psi_0$ changes from a single-peaked WP at $t=0$ to a doubled peaked one 
at $t=T$,
namely a splitting of a WP occurs.
The final time of the fast-forward $T_F$ is related to $T$
as $T_F=\ep T/\bar{v}$.
$\theta$ is obtained by numerical integration of equation 
(\ref{etheta2}) under the 
boundary conditions $\frac{\pa \theta}{\pa x}(x=0) = 0$ and $\theta (x=0)=0$.
The spatio-temporal dependence of $\theta$ is shown in figure \ref{theta4}.

\begin{figure}[h]
\begin{center}
\includegraphics[width=7cm]{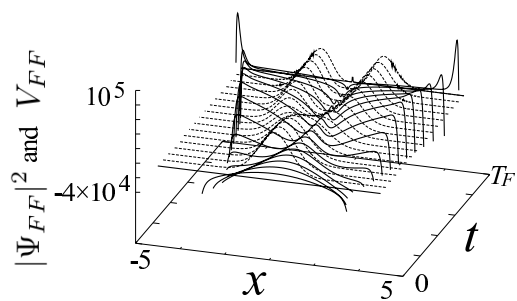}
\end{center}
\caption{\label{fig:epsart}
Spatio-temporal dependence of $|\Psi_{FF}|^2\times 10^5$ (broken line)
and $V_{FF}$ (solid line).  
The parameters are the same as in figure \ref{theta4}.
}
\label{VFF_WF_split}
\end{figure}

\begin{figure}[h]
\begin{center}
\includegraphics[width=7cm]{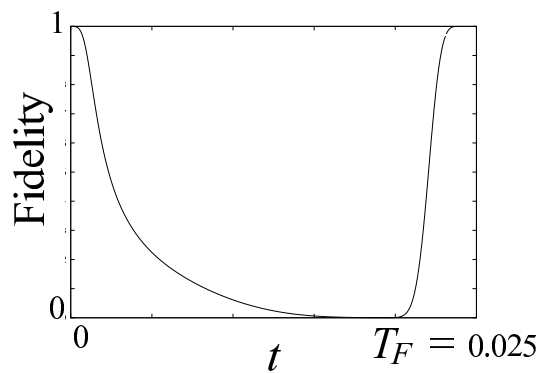}
\end{center}
\caption{\label{fig:epsart}
Time dependence of fidelity.  
}
\label{fidelity_split}
\end{figure}

The driving potential is calculated from equation (\ref{Vff0}) with use of 
equations (\ref{psi0_split}) and (\ref{V0_split}) and $\theta$.
$V_{FF}$ (solid line) and $|\Psi_{FF}|^2$ (broken line) 
which is accelerated by $V_{FF}$ are shown in 
figure \ref{VFF_WF_split}.
The conversion from a hump to a hollow in the central region of the potential
in figure \ref{VFF_WF_split} is caused by the deceleration of $\alpha(t)\ep$
which suppresses the splitting force. 
The fidelity defined by equation (\ref{fidelity}) 
is confirmed to be back to unity with a high numerical precision
($0.999$) at $T_F$ (see figure \ref{fidelity_split}).

\subsection{Fast-forward of wave packet squeezing: case of $\tilde{V}=0$}
\label{Fast-forward of wave packet squeezing: case of}
\begin{figure}[h]
\begin{center}
\includegraphics[width=7cm]{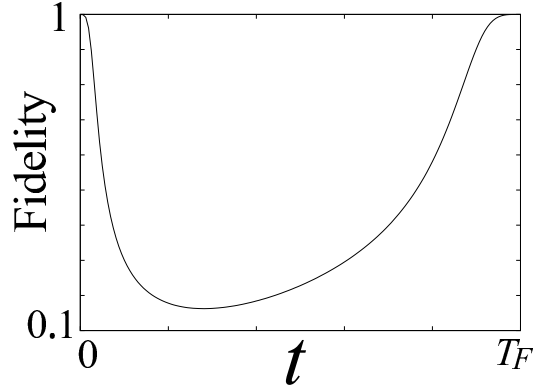}
\end{center}
\caption{\label{fig:epsart}
Time dependence of fidelity.  
}
\label{fidelity_B=0}
\end{figure}

The fast-forward of WP squeezing is also an important example of
the accelerated WF manipulation.
Let us consider a WF with a Gaussian distribution in a harmonic potential in 
1-dimension which is squeezed adiabatically.
The regularized standard state is represented as
\begin{eqnarray}
\Psi_0^{(reg)} = \phi_ne^{i\ep\theta} \equiv (\frac{m_0\omega}{\pi\hbar})^{1/4}
\exp[-\frac{m_0\omega}{2\hbar}x^2]
e^{i\ep\theta}.
\label{psisq0}
\end{eqnarray}
The regularized standard potential is 
\begin{eqnarray}
V_0^{(reg)} = 
\frac{m_0\omega^2}{2}x^2.
\label{v0sq0}
\end{eqnarray}
In the adiabatic dynamics, the potential curvature $\omega$ 
is gradually increased as
\begin{eqnarray}
\omega(t) = \omega_0 + \ep t,
\end{eqnarray}
where $\omega_0$ is a constant.
In this case, we can easily confirm that 
\begin{eqnarray}
\theta = -\frac{m_0}{4\hbar\omega}x^2
\label{theta50}
\end{eqnarray}
satisfies equation (\ref{etheta2}), and that $\tilde{V}$ in the regularized 
standard potential in equation (\ref{V'3})
vanishes, because
$\mbox{Im}(\frac{\pa\phi_n}{\pa \omega}/\phi_n)
=\mbox{Im}(\frac{\pa\phi_n}{\pa x}/\phi_n)=0$. 
With use of equations (\ref{psisq0}), (\ref{v0sq0}) and (\ref{theta50}) in
equation (\ref{Vff0}) we obtain the driving potential as 
\begin{eqnarray}
V_{FF} = [\frac{m_0\omega^2(\Lambda(t))}{2} -
\frac{d\alpha}{dt}\ep\frac{m_0}{4\omega(\Lambda(t))}
-\frac{3m_0}{8\omega^2(\Lambda(t))}\alpha^2\ep^2]x^2,
\label{vffexp}
\end{eqnarray}
where $\Lambda(t)$ is defined by equation (\ref{lam}).
\begin{figure}[h]
\begin{center}
\includegraphics[width=10cm]{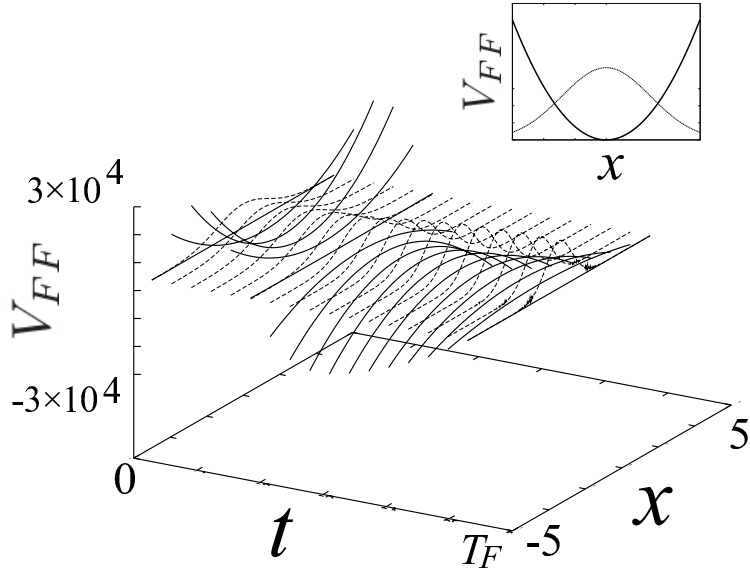}
\end{center}
\caption{\label{fig:epsart}
Spatio-temporal dependence of the driving potential (solid line) and
the amplitude of the fast-forwarded state $|\Psi_{FF}|^2\times 3.0\times 10^4$ 
(broken line).
The parameters are taken as $\omega_0=0.25$, $\ep T = 0.75$, 
$T_{F} = 0.025$ and $\frac{\hbar}{m_0}=1.0$.
The inset represents the potential (thick line) and wave function (thin line)  
distribution for $-3\le x \le 3$ at $t=0$.
}
\label{VFF_B=0}
\end{figure}
Spatio-temporal dependence of the driving potential $V_{FF}$ 
is shown in figure \ref{VFF_B=0}.
The conversion of the potential curvature from positive to negative in 
figure \ref{VFF_B=0} is caused by the deceleration of $\alpha(t)\ep$
which suppresses the squeezing force.
Time dependence of the fidelity is shown in figure \ref{fidelity_B=0}.
The fidelity becomes unity with a numerical precision of $0.999$ 
at the final time of the 
fast-forward and the WP squeezing has been carried out in a short time. 

\subsection{Fast-forward of wave packet squeezing: case of $\tilde{V}\ne 0$}
\begin{figure}[h]
\begin{center}
\includegraphics[width=8cm]{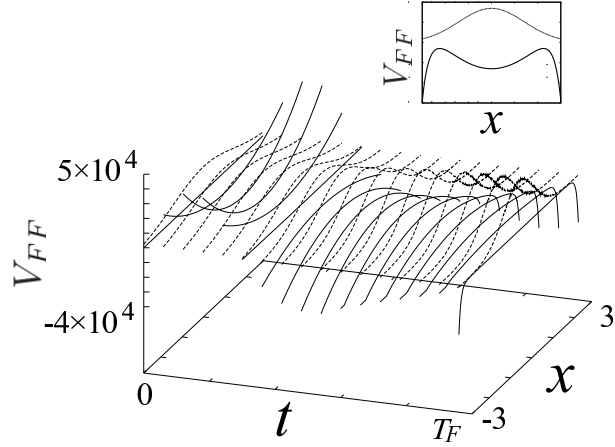}
\end{center}
\caption{\label{fig:epsart}
Spatio-temporal dependence of $V_{FF}$ (solid lines).
The broken lines represent the corresponding WF 
$|\Psi_{FF}|^2\times 6.0 \times 10^4$.
The parameters are taken as $\omega_0=0.25$, $\ep T = 0.75$, 
$T_{F} = 0.025$, $\frac{\hbar}{m_0}=1.0$ and $B = 0.1$.
The inset represents the distribution of the potential (thick line) and WF 
(thin line) at $t=0$.
}
\label{VFF_squeeze}
\end{figure}

In the previous examples, the right-hand side of equation (\ref{V'3}) was 
always vanishing
(for examples because of the absence of  
space-dependent phase of $O(1)$ in the standard WF). 
While in many practical cases we do not have to choose
a stationary state with a space-dependent phase of $O(1)$ for a standard state,
WF can in principle have such a phase.
Let us finally show the fast-forward of WP-squeezing as an example of the
fast-forward with non-zero $\tilde{V}$ 
by having recourse to
a stationary state
with space-dependent phase $\eta$ of $O(1)$.
Here the regularized standard state with $E_n=0$ 
is chosen as
\begin{eqnarray}
\Psi_0^{(reg)} = \phi_ne^{i\ep\theta} \equiv (\frac{m_0\omega}{\pi\hbar})^{1/4}
\exp[-\frac{m_0\omega}{2\hbar}x^2]e^{i\eta}
e^{i\ep\theta},
\label{psisq}
\end{eqnarray}
with $\eta$ given as
\begin{eqnarray}
\eta = B\int_0^x \exp[\frac{m_0\omega}{\hbar} x'^{2}] dx',
\label{eta2}
\end{eqnarray}
where $B$ is a real constant.
The regularized standard potential 
that guarantees the eigenstate in equation (\ref{psisq})
is written as
\begin{eqnarray}
V_0^{(reg)} = -\frac{\hbar\omega}{2} + \frac{m_0\omega^2}{2}x^2
-\frac{\hbar^2B^2}{2m_0}\exp[\frac{2m_0\omega}{\hbar}x^2] + \ep\tilde{V}.
\label{v0sq}
\end{eqnarray}
$\omega$ is a parameter which is adiabatically changed as
\begin{eqnarray}
\omega(t) = \omega_0 + \ep t,
\end{eqnarray}
where $\omega_0$ is the initial value of $\omega$.
$\Psi_0^{(reg)}$ with $\theta = 0$
and constant $\omega$  in equation (\ref{psisq}) stands for the zero-energy
eigenstate trapped in the central hollow of the potential barrier
$V_0^{(reg)}$ with $\tilde{V}=0$ (see the inset in figure \ref{VFF_squeeze}).
And the WP is squeezed adiabatically. 

\begin{figure}[h]
\begin{center}
\includegraphics[width=9cm]{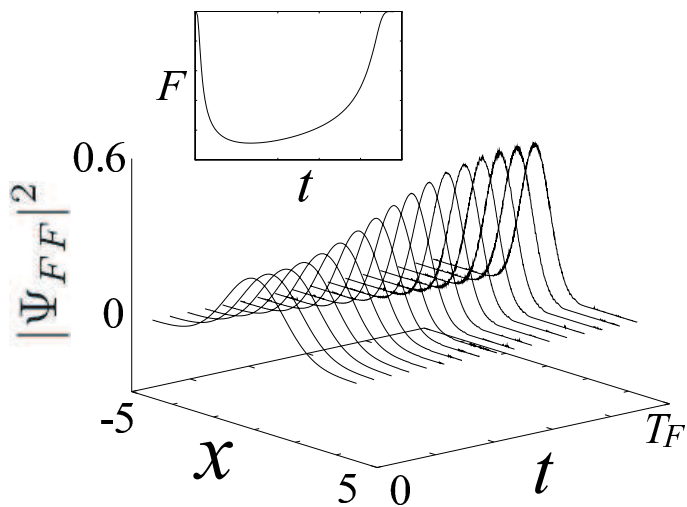}
\end{center}
\caption{\label{fig:epsart}
Spatio-temporal dependence of $|\Psi_{FF}|^2\times 10^5$. 
The inset represents the time dependence of the fidelity for $0\le t\le T_F$.
The parameters are the same as in figure \ref{VFF_squeeze}. 
}
\label{WF_squeeze}
\end{figure}

From equation (\ref{etheta2}), which is not affected by $\eta$, 
$\theta$ is obtained as
\begin{eqnarray}
\theta = -\frac{m_0}{4\hbar\omega}x^2.
\label{theta5}
\end{eqnarray}
Substituting equations (\ref{eta2}) and (\ref{theta5}) and $\phi_n$ in 
equation (\ref{psisq})
into equation (\ref{V'3}), $\tilde{V}$ is obtained as
\begin{eqnarray}
\tilde{V} = -m_0B\int_0^xx'^2\exp[\frac{m_0\omega}{\hbar}x'^2]dx'
+\frac{\hbar}{2\omega}Bx\exp[\frac{m_0\omega}{\hbar}x^2].
\label{vtilde}
\end{eqnarray}
Substituting 
equations (\ref{theta5}) and (\ref{vtilde})
into equation (\ref{Vff0}), we obtain the driving potential $V_{FF}$,
whose spatio-temporal dependence is shown in 
figure \ref{VFF_squeeze}.
The solid and broken lines represent the driving potential and 
the corresponding amplitude of WF, respectively.
The conversion of the potential curvature from positive to negative in 
figure \ref{VFF_squeeze} 
can be explained in the same way as in the case of the example 
(\ref{Fast-forward of wave packet squeezing: case of}).
The amplitude of fast-forwarded WF is shown in 
figure \ref{WF_squeeze}.
The inset in figure \ref{WF_squeeze} 
represents the time dependence of the fidelity
for $0\le t\le T_F$.
We confirmed that the fidelity 
comes back to unity with a high precision at $T_{F}$.

\section{Conclusion}
\label{Conclusion}
We have shown the way to accelerate the adiabatic dynamics
in microscopic and macroscopic quantum mechanics, 
with
use of infinitely-large 
magnification factor of the fast-forward and regularization of
standard states and Hamiltonian.
One can obtain the target state 
exactly except for spatially-uniform (dynamical and adiabatic) phases 
in any desired short time,
while in the adiabatic dynamics 
the target state is accessible after infinite
time.
A noble feature of the fast-forward of adiabatic dynamics,
which is distinct from our previous theory,
is that
we should regularize in advance the standard state so as to satisfy 
Schr$\ddot{\mbox{o}}$dinger equation up to $O(\ep)$ 
for a small but finite rate $\ep$
of temporal change in Hamiltonian.
To avoid the divergence of the driving potential,
we used a new form of the additional phase $f$
in equation (\ref{f1}) which differs from one in our previous
work (Masuda \& Nakamura 2008).

Typical examples of the fast-forward of adiabatic dynamics were shown, e.g., 
the wave function transport, and the wave packet splitting and squeezing.
The fast-forward of adiabatic transport makes the ideal transport of wave 
functions possible. The wave packet is rapidly transported to the 
adiabatically accessible targeted position,
with leaving neither disturbance nor 
oscillation after
the transport: 
the wave function becomes stationary again
in the confining potential at the end of the fast-forward.  
We confirmed that the fidelity first decreases from unity due to the phase $f$
of the fast-forwarded state, but at the final time it becomes unity with 
high precision again. 
The examples of the wave packet squeezing and splitting show
a way to fast-forward the adiabatic manipulation of wave packets
without leaving any residual disturbance.
The framework of this theory is applicable to the macroscopic quantum mechanics
described by the nonlinear Schr$\ddot{\mbox{o}}$dinger equation,
and one can expect 
the high-speed manipulations of wave packets in Bose Einstein condensates,
which will be reported elsewhere.


\begin{acknowledgements}
S.M. thanks Japan Society for the Promotion 
of Science for its financial support.
K.N. is grateful for the support received through a project of the Uzbek Academy
of Sciences (FA-F2-084).
We thanks 
D. Matraslov, B. Baizakov, B. Abdullaev, M. Musakhanov,
S. Tanimura and S. Sawada for useful discussions.
\end{acknowledgements}

\appendix{Derivation of equation (2.18) from continuity equation}
\label{Approach from the continuity equation}
Equation (\ref{etheta2}) is also obtained from the continuity equation.
Suppose that Schr$\ddot{\mbox{o}}$dinger equation for the
regularized standard state $\Psi_0^{(reg)}$ is given by equation (\ref{schr1_1}).
And suppose $\Psi_0^{(reg)}$ to be represented by equation (\ref{standard1}),
where $\phi_n$ satisfies equations (\ref{dphidt}) and (\ref{tis}).
Then we have 
\begin{eqnarray}
\frac{d}{dt}|\Psi_0^{(reg)}|^2=\frac{d\Psi_0^{(reg)\ast}}{dt}\Psi_0^{(reg)} + 
\Psi_0^{(reg)\ast}\frac{d\Psi_0^{(reg)}}{dt} = 
2\varepsilon\mbox{Re}[\frac{\pa\phi_n^\ast}{\pa R}\phi_n],
\label{dp2dt0}
\end{eqnarray}
where we used equation (\ref{dphidt}).
On the other hand, we have the continuity equation:
\begin{eqnarray}
\frac{d}{dt}|\Psi_0^{(reg)}|^2=-\frac{\hbar}{m_0}\nab\cdot\mbox{Im}[
\Psi_0^{(reg)\ast}\nab\Psi_0^{(reg)}].\label{dp2dt1}
\end{eqnarray}
Substituting equation (\ref{standard1}) into equation (\ref{dp2dt1}),
we have
\begin{eqnarray}
\frac{d}{dt}|\Psi_0^{(reg)}|^2
= -\frac{\hbar}{m_0}\nab\cdot\mbox{Im}[\phi_n^\ast\nab\phi_n
+ i\ep|\phi_n|^2\nab\theta]
.\label{im}
\end{eqnarray}
By the way we obtain
\begin{eqnarray}
\nab\cdot\mbox{Im}[\phi_n^\ast\nab\phi_n]=0
\label{im0}
\end{eqnarray}
by multiplying $\phi_n^\ast$ on the both sides of equation 
(\ref{tis}) and taking
their imaginary part. 
Therefore by using equations (\ref{dp2dt0}), (\ref{im}) and (\ref{im0}), 
we obtain
\begin{eqnarray}
\frac{\hbar}{m_0}\nab\cdot[|\phi_n|^2\nab\theta]
=2\mbox{Re}[\frac{\pa\phi_n^\ast}{\pa R}\phi_n].\label{etheta0}
\end{eqnarray}
$\nab\theta$ must satisfy this equation.
From equation (\ref{etheta0}), we can reach equation (\ref{etheta2}).

\appendix{Additional phase of fast-forwarded state}
\label{Additional phase for fast-forward state}
We show that the gradient of the additional phase $f$ in equation (\ref{nabf})
satisfies equation (\ref{feq44_2}).
$\Psi_0^{(reg)}$ in
equation (\ref{standard1}) includes the amplitude factor
\begin{eqnarray}
\phi_n(\largex,R(t)) = |\phi_n|e^{i\eta(\largex,R(t))},\label{phin}
\end{eqnarray}
where $\eta=\eta(\largex,R(t))$ is the space-dependent phase.
With the use of equations (\ref{tis}) and (\ref{phin}) we have
\begin{eqnarray}
\phi_n^\ast\nab^2\phi_n-\phi_n\nab^2\phi_n^\ast = 
2|\phi_n|[2\nab\eta\cdot\nab|\phi_n|
+(\nab^2\eta)|\phi_n|]
=0.
\end{eqnarray}
Thus, $\eta$ satisfies
\begin{eqnarray}
2\nab|\phi_n|\nab\eta + |\phi_n|\nab^2\eta=0.
\label{eta1}
\end{eqnarray}
Noting $\ps$ in equation (\ref{feq44_2}) is now given by
$\ps (t) = \Psi_0^{(reg)}(\lam)$ 
in the regularized case,
we substitute equations (\ref{standard1}) and (\ref{phin}) into 
equation (\ref{feq44_2}),
and obtain
\begin{eqnarray}
|\phi_n|^2\nab^2f+2|\phi_n|\nab|\phi_n|\cdot \nab f
&+& (\alpha-1)[-2|\phi_n|\nab(\eta+\ep\theta)\cdot\nab|\phi_n|\nonumber\\
&& -|\phi_n|^2\nab^2(\eta+\ep\theta)] = 0.
\label{add_2}
\end{eqnarray}
By using equation (\ref{eta1}) in equation (\ref{add_2}), 
we can easily check that
$\nab f$ defined by equation (\ref{nabf}) satisfies equation (\ref{add_2}), 
i.e., equation (\ref{feq44_2}).



\label{lastpage}
\end{document}